\documentstyle[aps,twocolumn,floats,prl]{revtex}
\input epsf

\newcommand\lsim{\mathrel{\rlap{\lower4pt\hbox{\hskip1pt$\sim$}}
        \raise1pt\hbox{$<$}}}
\newcommand\gsim{\mathrel{\rlap{\lower4pt\hbox{\hskip1pt$\sim$}}
        \raise1pt\hbox{$>$}}}

\begin{document}

\twocolumn[\hsize\textwidth\columnwidth\hsize\csname
@twocolumnfalse\endcsname

\title{Unitarity Bounds and the Cuspy Halo Problem}

\author{Lam Hui}

\address{
Department of Physics, 
Columbia University, 538 West 120th Street, New York, NY 10027\\
Institute for Advanced Study, School of Natural Sciences, 
Princeton, NJ 08540\\ {\tt lhui@astro.columbia.edu}}
\maketitle
\begin{abstract}
Conventional Cold Dark Matter cosmological models predict small
scale structures, such as cuspy 
halos, which are in apparent conflict with observations. 
Several alternative scenarios based on modifying fundamental properties 
of the dark matter have been proposed.
We show that general principles of quantum mechanics, in particular unitarity,
imply interesting constraints on two 
proposals: collisional dark matter proposed by Spergel \& Steinhardt, and 
strongly annihilating dark matter proposed by Kaplinghat, Knox \& Turner.
Efficient scattering required in both implies
$m \lsim 12$ GeV and $m \lsim 25$ GeV respectively. The same arguments 
show that the strong annihilation in
the second scenario implies the presence of significant
elastic scattering, particularly for large enough masses.
Recently, a variant of the collisional scenario has been
advocated to satisfy simultaneously constraints from dwarf
galaxies to clusters, with a cross section that scales inversely
with velocity. We show that this scenario likely involves
super-elastic processes, and the associated kinetic energy change
must be taken into account when making predictions.
Exceptions and implications for experimental searches are discussed.
\vskip 0.3truecm
98.89.Cq; 98.80.Es; 98.62.Gq; 11.80.Et; 11.55Bq
\end{abstract}
\vskip 0.5truecm
]


\section{Introduction}
\label{intro}

There is a long history of efforts to constrain dark matter
properties from galactic structure (e.g. \cite{gunntremaine}). 
Recent numerical simulations \cite{simulations}
sharpen the predictions of Cold Dark Matter (CDM) structure 
formation models, and apparent discrepancies with the
observed properties of structures from galactic to cluster scales are 
uncovered. The main one that has attracted a lot of attention is the 
cuspy halo problem , namely that CDM models predict halos
that have a high density core or have an inner profile that
is too steep compared to observations (\cite{earlywork}, but
see also \cite{vanDB}). This has encouraged several proposals that dark 
matter might have properties different from those of conventional CDM
(see \cite{dave} and summary therein). 

On the other hand, general principles of 
quantum mechanics impose non-trivial constraints on some of these models.
We focus here on the proposals of collisional or strongly 
self-interacting dark matter (SIDM)
by \cite{ss} and of strongly annihilating dark matter (SADM) by
\cite{kkt}. Both require high level of interaction by particle physics 
standard: an elastic scattering cross-section of 
$\sigma_{\rm el.} \sim 10^{-24} (m_X/{\, \rm GeV})
{\, \rm cm^2}$ for
the former and an annihilation cross-section of
$\sigma_{\rm ann.} v_{\rm rel} \sim
10^{-28} (m_X/{\, \rm GeV}) {\, \rm cm}^2$
for the latter, where $m_X$ is the
particle mass, and $v_{\rm rel}$ is the relative velocity of approach.
The proposed dark matter is therefore quite different from usual
candidates such as the axion or neutralino. We show that
the unitarity of the scattering matrix, together with a few reasonable 
assumptions, imposes interesting particle mass bounds 
as well as other physical constraints. This is done while making
minimal assumptions about the nature of the interactions.
Our results complement constraints from experiments
or astrophysical considerations e.g. \cite{ben00}.

Griest and Kamionkowski \cite{gk} previously derived similar mass
bounds related to the freeze-out density of thermal relics, assuming
2-body final states. In \S \ref{optical}, we provide a general derivation
for arbitrary final states using the classic optical theorem \cite{OT}.
We summarize our findings in \S \ref{discuss}, discuss exceptions to
our bounds, and other solutions to the
cuspy halo problem.

\section{Deriving the Unitarity Bounds}
\label{optical}

Different versions of the unitarity bounds can be found in
many textbooks, and can be most easily understood using
non-relativistic quantum mechanics (e.g. \cite{landau}),
which is probably adequate for our purpose.
However, the results and derivation given here might be
of wider interest e.g. for estimating thermal relic density.
Here we follow closely the field theory treatment of \cite{weinberg}.

The optical theorem \cite{OT} is a powerful consequence of
the unitarity of the scattering matrix $S$, 
i.e. $S^\dagger S = 1$, which implies
$(1-S)^\dagger (1-S) = (1-S^\dagger) + (1-S)$, or 
\begin{eqnarray}
\label{begin}
\int d\gamma \langle \beta |1-S| \gamma \rangle 
\langle \gamma |1-S^\dagger| \alpha \rangle
= 
2 {\, \rm Re \, }\langle \beta |1-S| \alpha \rangle
\end{eqnarray}
where $\alpha$ and $\beta$ represent two specified states
and $\gamma$ represents a complete set of states with
measure $d \gamma$. Using the definition of the 
scattering amplitude $A_{\beta\alpha}$ 
\begin{equation}
\langle \beta | 1 - S | \alpha \rangle \equiv
 -i (2\pi)^4 \delta^4 (p_\beta
- p_\alpha) A_{\beta\alpha}
\label{amplitude}
\end{equation}
where $p_\beta$ and $p_\alpha$ are the total four-momenta, 
one obtains
\begin{equation}
\int d\gamma (2\pi)^4 \delta^4 (p_\alpha - p_\gamma) |A_{\gamma\alpha}|^2
= 2 \, {\rm Im \,} A_{\alpha\alpha}
\label{OT1}
\end{equation}
if $\beta = \alpha$ in eq. (\ref{begin}). 
We are interested in the case where $\alpha$ represents a 2-body state
of $X + X$ or $X + \bar X$ approaching each other.
The final state $\gamma$, on the other hand, is completely general, and
the integration over $\gamma$ covers the entire spectrum
of possible final states.
To be more precise,
suppose $| \alpha \rangle = | k_1, s^z_1; k_2, s^z_2; n\rangle$
where $s^z_1$ and $s^z_2$ represent the spin-states of
the two incoming particles with
spins $s_1$ and $s_2$ (in our particular case,
$s_1 = s_2$) while $k_1$ and $k_2$ are their
respective 4-momenta, and $n$ labels the particle
types (e.g. mass, etc.).
Recalling that $d\sigma/d\gamma \propto |A_{\gamma\alpha}|^2$,
eq. (\ref{OT1}) gives in the center of mass frame (adopted
hereafter i.e. ${\bf k_1} +{\bf k_2}=0$):
\begin{equation}
\int d\gamma {d\sigma \over d\gamma}
(\alpha \rightarrow \gamma)
= {{\, \rm Im \,} A_{\alpha\alpha} \over 2(E_1 + E_2) |{\bf k_1}|}
\label{OT2}
\end{equation}
where the left hand side is exactly the total cross-section.
This is the optical theorem.
It states that the total cross-section for scattering from
a two-body initial state {\it to all possible} final states
equals the imaginary part of the two-body {\it to two-body}
forward scattering amplitude.

To use this theorem, we expand
the scattering amplitude in terms of partial waves
i.e. states labeled as $|{\bf k_{\rm tot.}}, E_{\rm tot.}, 
j, j^z, \ell, s, n\rangle$ where ${\bf k_{\rm tot.}}$ is
the total linear momentum ($=0$ in the center
of mass frame), $E_{\rm tot.}$ is the total energy 
, 
$j$ is the total angular momentum, $j^z$ is its z-component,
$\ell$ is the orbital momentum and $s$ is the total spin.
Inserting appropriate complete sets of partial wave outer products into
eq. (\ref{amplitude}), we obtain
\begin{eqnarray}
\label{Abetaalpha}
A_{\beta\alpha} && = 4 i (2\pi)^2 
[E_1' + E_2' / |{\bf k_1'}|]^{{1\over 2}}
 [E_1 + E_2 / |{\bf k_1}|]^{{1\over 2}}
\\ \nonumber &&
\sum_{j,j^z} \sum_{\ell',s',\ell,s} 
\langle \ell' s' n'|1-S| \ell s n\rangle_{j,E_{\rm tot.}}
\\ \nonumber &&
\sum_{{\ell^z}',{s^z}'}
\langle s' {s^z}' | {s_1^z}' {s_2^z}' \rangle_{s_1', s_2'}
\langle j j^z | {\ell^z}' {s^z}'\rangle_{\ell',s'} 
\langle {\bf \hat k_1'} | \ell' {\ell^z}' \rangle
\\ \nonumber &&
\sum_{{\ell^z},{s^z}}
\langle s {s^z} | {s_1^z} {s_2^z} \rangle_{s_1, s_2}^*
\langle j j^z | {\ell^z} {s^z}\rangle_{\ell,s}^*
\langle {\bf \hat k_1} | \ell {\ell^z} \rangle^*
\end{eqnarray}
where the crucial assumption is that $S$ is rotationally
invariant and so $j$ and $j_z$ are conserved, in addition
to energy conserving.
The notation $\langle \ell' s' n'|1-S| \ell s n\rangle_{j,E_{\rm tot.}}$
emphasizes that $S$ is diagonal in $j$, $j^z$ and $E_{\rm tot}$
but the $j^z$ dependence drops out because 
$S$ commutes
with $J_x \pm i J_y$.
The inner products $\langle s s^z | s^z_1 s^z_2 
\rangle_{s_1,s_2}$ and $\langle j j^z | \ell^z s^z \rangle_{\ell, s}$
give the Clebsch-Gordon coefficients, and
$\langle {\bf \hat k_1} |
\ell \ell^z \rangle = Y_{\ell \ell^z} ({\bf \hat k_1})$ is
the spherical harmonic function. 
We assume ${\bf \hat k_1} = {\bf \hat z}$ in which
case $Y_{\ell \ell^z} ({\bf \hat k_1}) = \delta_{\ell^z, 0}
\sqrt{2\ell + 1/(4\pi)}$. The index $\beta$
denotes a 2-body final state $|k_1', s_1'; k_2', s_2'; n'\rangle$.

Setting $\beta = \alpha$, and averaging over the spin-states
(i.e. $(2 s_1 + 1)^{-1}
(2 s_2 + 1)^{-1} \sum_{s_1^z, s_2^z}$) on both sides of
eq. (\ref{OT2}), the optical theorem, we obtain \cite{weinberg}:
\begin{eqnarray}
\label{OTpartial}
\sigma_{\rm tot.} 
& = &
{2\pi \over {|{\bf k_1}|^2 (2s_1 + 1) (2s_2 + 1)}}
\sum_j (2j+1) 
\\ \nonumber && 
\sum_{\ell,s}
{\, \rm Re\,} 
\langle \ell s n | 1 - S |  \ell s n\rangle_{j, E_{\rm tot.}}
\end{eqnarray}
This gives the total spin-averaged cross-section for
scattering from $X + X$ or $X + \bar X$ to all possible final states. 

For $X + \bar X$ annihilation, we exclude from the above
the contribution due to elastic scattering (where type
and mass of particles do not change i.e. $X + \bar X
\rightarrow X + \bar X$, implying $|{\bf k_1'}| = |{\bf k_1}|$)
\cite{gk}.
To do so, we need the following expression for 
2-body to 2-body scattering cross-section:
\begin{equation}
\label{Asigma2}
{d\sigma\over d\beta} d\beta
= {|A_{\beta\alpha}|^2 \over 4 (E_1 + E_2) |{\bf k_1}|}
(2\pi)^4 \delta^4 (p_\beta - p_\alpha) d\beta \, ,
\end{equation}
We average over initial spin states and
integrate over outgoing momenta, but focus on the
elastic contribution ($n'$
in $|\beta\rangle = |k_1',s_1';k_2',s_2';n'\rangle$
is set to $n$ in $|\alpha\rangle$) \cite{weinberg}:
\begin{eqnarray}
\label{elastic}
\sigma_{\rm el.} 
&=& {\pi \over |{\bf k_1}|^2 (2s_1 + 1) (2s_2 + 1)} \sum_j (2j+1)
\\ \nonumber &&
\sum_{\ell,s,\ell',s'} |\langle \ell' s' n| 1 - S | \ell s n
\rangle_{j,E_{\rm tot.}}|^2 
\end{eqnarray}
The above is the total cross-section for elastic scattering 
(note: the same expression also describes $X + X \rightarrow
X + X$ elastic scattering) that
has to be subtracted from $\sigma_{\rm tot.}$ to yield
the total inelastic
scattering cross-section, which is relevant for
annihilation into all possible final states:
\begin{eqnarray}
\label{OTfinal}
&& \sigma_{\rm inel.} =
{\pi 
\over k_1^2 (2s_1 + 1) (2s_2 + 1)} 
\sum_j (2j+1) \\ \nonumber && 
\sum_{\ell,s} [1 - 
|\langle \ell s n| S | \ell s n\rangle |^2 
- \sum_{\ell'\ne\ell,s'\ne s}
|\langle \ell' s' n| 1 - S | \ell s n\rangle |^2]
\end{eqnarray}
From eq. (\ref{OTpartial}) \& (\ref{OTfinal}), we can
derive two bounds:
\begin{eqnarray}
\label{stot}
&& \sigma_{\rm tot.} \le 4\pi [|{\bf k_1}|^2
(2s_1 + 1)(2s_2+1)]^{-1} \sum_j \sum_{\ell,s}2j+1\\ 
\label{sinel}
&& \sigma_{\rm inel.} \le \pi [|{\bf k_1}|^2
(2s_1 + 1)(2s_2+1)]^{-1} \sum_j \sum_{\ell,s}2j+1
\end{eqnarray}
The first inequality
uses $|\langle \ell s n | S | \ell s n\rangle|^2 \le 1$, obtained
from $\int d\gamma \langle \ell s n | S^\dagger
|\gamma\rangle\langle\gamma|S|\ell s n\rangle \ge
|\langle \ell s n |S| \ell s n\rangle|^2$ and
$S^\dagger S = 1$. A similar bound can be derived
for $\sigma_{\rm el.}$ as well, which coincides
exactly with that for $\sigma_{\rm tot.}$. 

We pause to note that the above bounds assume only
unitarity and the conservation of total energy and
linear and angular momentum. No assumptions are
made about the nature of the particles, whether they
are composite or point-like. Nor do we assume 
the number of particles in the final states.
To obtain useful limits from the bounds, we take
the low velocity limit. Assuming the scattering
amplitude $A_{\beta\alpha}$ is an analytic function of
${\bf k_1}$ as ${\bf k_1} \rightarrow 0$ (exceptions will
be discussed in \S \ref{discuss}), and noting that
$k^\ell \langle {\bf \hat k} | \ell \ell^z \rangle$
is a polynomial function of ${\bf k}$, we expect the
$\ell$ partial wave contribution to 
$A_{\beta\alpha}$ (eq. \ref{Abetaalpha})
to scale as $|{\bf k_1}|^\ell$. This means
in the low velocity limit, as is relevant
for our purpose (typical velocity dispersion in halos
range from $10$ to $1000$ km/s $\ll c$), the $\ell = 0$ or
s-wave contribution dominates. Setting $\ell=0$ in eq.
(\ref{stot}), (\ref{sinel}):
\begin{eqnarray}
\sigma_{\rm tot.} \le 16\pi/(m_X v_{\rm rel})^2
\quad , \quad
\sigma_{\rm inel.} v_{\rm rel} \le 4\pi/(m_X^2 v_{\rm rel})
\end{eqnarray}
where $k_1^2 = m_X^2 |{\bf v_2}-{\bf v_1}|^2/4 =
m_X^2 v_{\rm rel}^2/4$ is used. 
The second inequality agrees with \cite{gk}.
Hence,
\begin{eqnarray}
\label{stotF}
&& \sigma_{\rm tot.} \le 1.76 \times 10^{-17} {\, \rm cm^2}
\left[{{\rm GeV} \over m_X} \right]^2 \left[{10{\,\rm km \, s^{-1}} \over v_{\rm rel.}} \right] ^2 \\
\label{sinelF}
&& \sigma_{\rm inel.} v_{\rm rel.} \le  1.5 \times 10^{-22} {\,\rm cm^2}
\left[{\rm GeV} \over m_X \right]^2 \left[10{\,\rm km \, s^{-1}} \over 
v_{\rm rel.} \right] 
\end{eqnarray}
Furthermore, if $\sigma_{\rm inel.}$ is bounded from below, say 
$\sigma_{\rm inel.} \ge 
\sigma_{\rm ann.}$, one
can derive a lower bound on $\sigma_{\rm el.}$ using
eq. (\ref{elastic}) \& (\ref{OTfinal}), and setting
$\ell = 0$. Defining
$\langle X \rangle_J \equiv
[(2s_1+1)(2s_2+1)]^{-1}\sum_{j,\ell,s} (2j+1) X$, using
$S$ at the moment to denote $\langle \ell sn|S|\ell sn\rangle$, 
and noting that $\langle 1 \rangle_J = 1$ for $\ell=0$, 
it can be shown $(\pi/k_1^2) (1 - \langle |S| \rangle_J^2) \ge
(\pi/k_1^2) (1 - \langle |S|^2 \rangle_J)
\ge \sigma_{\rm inel.} \ge \sigma_{\rm ann.}$, which
implies $\langle |S| \rangle_J \le \sqrt{1-k_1^2\sigma_{\rm ann.}/\pi}$.
Also, $\sigma_{\rm el.} \ge (\pi/k_1^2) \langle |1 - S|^2 \rangle_J
\ge (\pi/k_1^2) \langle (1 - |S|)^2 \rangle_J
\ge (\pi/k_1^2) (1 - \langle |S| \rangle_J )^2$. Combining, we have
\begin{eqnarray}
\label{selF}
\sigma_{\rm el.} \ge (\pi/k_1^2) 
[\,1-\sqrt{1-k_1^2 \sigma_{\rm ann.}/\pi}\,]^2
\end{eqnarray}
This tells us that the elastic scattering cross-section cannot
be arbitrarily small given a non-vanishing inelastic cross-section,
e.g. via annihilation.

The above 3 bounds are the main results of this section.
Two more results will be useful for our later discussions.
For two-body to two-body processes, recall that 
the $\ell$, $\ell'$ contribution to $A_{\beta\alpha}$
scales as $|{\bf k_1}|^\ell |{\bf k_1'}|^{\ell'}$. Using 
$d\sigma/d\Omega = |A_{\beta\alpha}|^2 (|{\bf k_1'}|/|{\bf k_1}|)
/[64\pi^2(E_1+E_2)^2]$ (obtained from eq. \ref{Asigma2}
by integrating over $\beta$ except for solid angle $\Omega$),
it can be seen that for elastic scattering, where
$|{\bf k_1'}|=|{\bf k_1}|$, 
\begin{eqnarray}
\label{elasticV}
d\sigma/d\Omega 
\rightarrow {\, \rm const.} [1 + O(v_{\rm rel.})]
\end{eqnarray}
as $|{\bf k_1}| \rightarrow 0$.
For inelastic scattering where the system gains kinetic energy
by losing rest mass
(e.g. de-excitation of a composite particle or annihilation), 
since $|{\bf k_1'}|$ approaches a non-zero value
as $|{\bf k_1}| \rightarrow 0$, we have
\begin{eqnarray}
\label{inelasticV}
d\sigma/d\Omega \rightarrow ({\, \rm const.}/ v_{\rm rel.})
[1 + O(v_{\rm rel.})]
\end{eqnarray}
instead in the low velocity limit.
The opposite case where the particle gains mass is discussed in
\cite{weinberg}.

\section{Discussion}
\label{discuss}

We can derive the following four constraints for strongly self-interacting
dark matter (SIDM) \cite{ss} and strongly annihilating dark matter
(SADM) \cite{kkt}.


{\bf 1.} The range $\sigma_{\rm el.} \sim
10^{-24} - 10^{-23} {\, \rm cm^2} (m_X/{\, \rm GeV})$ is given by
\cite{dave} for SIDM to yield the desired halo properties.
Using the lower $\sigma_{\rm el.}$, and 
$v_{\rm rel.} \sim 1000$ km/s (appropriate for clusters),
eq. (\ref{stotF}) tells us $m_X \lsim 12$ GeV
for SIDM.

{\bf 2.} The annihilation cross-section from \cite{kkt},
$\sigma_{\rm ann.} v_{\rm rel.}
\sim 10^{-28} {\, \rm cm^2} (m_X/{\,\rm GeV})$, together with
eq. (\ref{sinelF}) and $v_{\rm rel.} \sim 1000$ km/s, 
gives us a bound of $m_X \lsim 25$ GeV for strongly
annihilating dark matter.

{\bf 3.} For SADM, efficient annihilation (a form
of inelastic scattering) inevitably
implies some elastic scattering as well.
From eq. (\ref{selF}), and using $v_{\rm vel} \sim 1000$ km/s
as before, we have
\begin{eqnarray}
\sigma_{\rm el.} & \ge & 4 \times 10^{-22} {\, \rm cm^2}
[{\, \rm GeV}/m_X]^2 \\ \nonumber &&
[1-\sqrt{1-7 \times 10^{-5} (m_X/{\, \rm GeV})^3}]^2
\end{eqnarray}
Two simple limiting cases: when $m_X$ is
close to the upper bound of $25$ GeV, $\sigma_{\rm el.} \gsim
4 \times 10^{-22} {\, \rm cm^2}$; when
$m_X$ is small, $\sigma_{\rm el.} \gsim 5 \times 10^{-31} 
{\,\rm cm^2} (m_X/{\, \rm
GeV})^4$. 
Hence, elastic scattering is inevitable in this scenario, but
can
be reduced by having a sufficiently small mass. 

{\bf 4.} Recent simulations suggest that the simplest version of SIDM
fails to match simultaneously the observed halo properties
from dwarf 
galaxies to clusters \cite{yoshida,dave}
(see also \cite{firmani}), which have $v_{\rm rel.}$
ranging over 3 orders of magnitude. It was suggested that
an {\it elastic} scattering cross-section of $\sigma \propto 1/v_{\rm rel.}$
might solve the problem. But as shown in eq. (\ref{elasticV}),
elastic scattering generally implies $\sigma \rightarrow$ constant
in the small velocity limit. Hence, $\sigma \propto 1/v_{\rm rel.}$
likely requires {\it inelastic} processes. As
eq. (\ref{inelasticV})
shows, processes in which the net kinetic energy increases
($|{\bf k_1'}| > |{\bf k_1}|$ in c.o.m. frame)
can give such a velocity dependence.
SADM provides an example. More generally,
the net kinetic energy increase (super-elasticity) must be taken into 
account when considering the viability of a model with 
$\sigma \propto 1/v_{\rm rel.}$
e.g. it may delay core collapse and make the core larger.
Note, however, the general considerations in the last section
does not forbid an elastic cross-section that
increases as $v_{\rm rel.}$ decreases e.g. the 
$O(v_{\rm rel.})$ term in eq. (\ref{elasticV})
can have a negative coefficient. A $1/v_{\rm rel.}$ power-law
may approximate such a cross-section, but
likely only for a limited range of $v_{\rm rel.}$. An example is
the neutron-neutron scattering cross-section, which approaches a constant 
for $|{\bf k_1}| \lsim 10^{-2}$ GeV, and scales
as $1/v_{\rm rel.}$ only for $10^{-2} \lsim |{\bf k_1}| \lsim 
5 \times 10^{-2}$ GeV \cite{blatt}.


It is helpful to mention here possible exceptions to the above
limits. Our bounds are obtained from eq. (\ref{stotF}) \&
(\ref{sinelF}), which are the $\ell = 0$ (s-wave) versions of
eq. (\ref{stot}) \& (\ref{sinel}). The argument for
putting $\ell = 0$ in the small velocity limit assumes
the analyticity of $A_{\beta\alpha}$ at ${\bf k_1}=0$.
The latter breaks down if the interaction is long-ranged,
e.g. Coulomb scattering. This is unlikely to be
relevant, because there are strong constraints on
dark matter with such long ranged interaction \cite{gould}. 
Our argument for the dominance of s-wave scattering
can also be invalid if there is a resonance. However, given
that the scattering cross section should vary smoothly over
three orders of magnitude in velocities
from dwarfs to clusters, a resonance seems 
unlikely. Finally, the most likely
situation in which the bounds break down is if
the particle has a large enough size, or the interaction has a 
large enough effective range, $R$, such that 
$|{\bf k_1}| R > 1$ (e.g. see \cite{pj}). 
In such cases, higher partial waves
in addition to s-wave generally contribute, and
$\sigma_{\rm tot.} \lsim 64 \pi R^2$ and our arguments turn
into a limit on $R$ \cite{gk}.
The condition $|{\bf k_1}| R > 1$ gives the most
stringent constraint on $R$ for $v_{\rm rel.} = 10$ km/s,
as appropriate for dwarf galaxies: $R \gsim 10^{-9} {\,\rm cm}
({\rm GeV}/m_X)$. One can compare this with $R$
for neutron-neutron scattering $\sim 10^{-13}$ cm \cite{blatt}.

It is intriguing that halo structure might be telling us 
the elementary properties, in particular the mass, of dark matter. 
It is interesting that several proposals to address the
cuspy halo problem, such as Warm Dark Matter \cite{wdm}
and Fuzzy Dark Matter \cite{fdm}
make explicit assumptions about the mass of the particles --
$m_X \sim 1$ keV and $m_X \sim 10^{-22}$ eV respectively.
For SIDM and SADM, astrophysical considerations generally only
put constraints on the cross-section per unit mass. We have shown
here that unitarity arguments imply a rather modest mass for
both scenarios as well. It is also
worth pointing out that our arguments,
with suitable modification to take into account bose enhancement and
multiple incoming particles,
can be extended to cover dark matter in the form of a bose condensate,
as has been proposed as yet another solution to the cuspy halo problem
\cite{fluidm}. They generally require small masses as well
$\lsim 10$ eV. 

A few issues are worth further investigation.
Wandelt et al. \cite{ben00} recently argued a version of
SIDM, where the dark matter interacts strongly also with baryons,
is experimentally viable, but requires $m_X \gsim 10^5$ GeV,
or $m_X \lsim 0.5$ GeV.
Our bound here is inconsistent with the large mass region
(but see exceptions above);
experimental constraints on the low mass region
will be very interesting ($\sigma_{\rm el.} \lsim 10^{-25} {\,\rm cm^2}$).
It would be useful to find a micro-physics
realization of the collisional scenario \cite{moha} or 
its variant where $\sigma$ scales appropriately with velocity
to match observations. The impact
of inelastic collisions on halo structures is worth
exploring in more detail.
It is also timely to reconsider possible
astrophysical solutions to the cuspy halo problem,
such as the use of mass loss mechanisms \cite{navarro}.
We hope to examine some of these issues
in the future.

\acknowledgments

The author thanks Manoj Kaplinghat, 
Avi Loeb, Dam Son, Paul Steinhardt and Igor Tkachev
for useful discussions, and Jonathan Feng,
Andrei Gruzinov and Yossi Nir 
for helpful comments.
Thanks are due to David Spergel for 
pointing out the example of neutron scattering,
and to Kim Griest and Marc Kamionkowski
for educating the author on elastic scattering, which
led to a tightening of a bound from an earlier version.
Support by the Taplin Fellowship at the IAS, and by
the Outstanding Junior Investigator Award from DOE, 
DE-FG02-92-ER40699, 
is gratefully acknowledged.

\end{document}